\documentclass[twocolumn,showpacs,preprintnumbers,amsmath,amssymb,superscriptaddress]{revtex4}

\usepackage{graphicx}% Include figure files
\usepackage{dcolumn}% Align table columns on decimal point
\usepackage{bm}% bold math

\begin{document}

\title{Universal Features of Information Spreading Efficiency on $d$-dimensional lattices}

\author{E. Agliari}
\affiliation{Dipartimento di Fisica, Universit\`a degli Studi di
Parma, viale Usberti 7/A, 43100 Parma, Italy}
\author{R. Burioni}
\affiliation{Dipartimento di Fisica, Universit\`a degli Studi di
Parma, viale Usberti 7/A, 43100 Parma, Italy}
\affiliation{INFN, Gruppo Collegato di Parma, viale Usberti 7/A, 43100 Parma, Italy}
\author{D. Cassi}
\affiliation{Dipartimento di Fisica, Universit\`a degli Studi di
Parma, viale Usberti 7/A, 43100 Parma, Italy}
\affiliation{INFN, Gruppo Collegato di Parma, viale Usberti 7/A, 43100 Parma, Italy}
\author{F.M. Neri}
\affiliation{Dipartimento di Fisica, Universit\`a degli Studi di
Parma, viale Usberti 7/A, 43100 Parma, Italy}

\begin{abstract}
A model for information spreading in a population of $N$ mobile
agents is extended to $d$-dimensional regular lattices. This model,
already studied on two-dimensional lattices, also takes into account
the degeneration of information as it passes from one agent to the
other. Here, we find that the structure of the underlying lattice
strongly affects the time $\tau$ at which the whole population has
been reached by information. By comparing numerical simulations with
mean-field calculations, we show that dimension $d=2$ is marginal
for this problem and mean-field calculations become exact for $d >
2$. Nevertheless, the striking nonmonotonic behavior exhibited by
the final degree of information with respect to $N$ and the lattice
size $L$ appears to be geometry independent.

\end{abstract}

 \maketitle

\section{\label{sec:intro}Introduction}

The problem of information spreading among a population has been
intensively studied in the last years and several aspects have been
focused upon \cite{earlier,gonzalez,huang,sole}.

The population is generally represented by means of a graph such
that an interaction (link) between two or more agents (nodes) means
that there is a flow of information among them. Recently, the
dynamics of agents making up the population has also been taken into
account \cite{earlier,gonzalez}. Not only does the mobility of
agents provide a realistic feature, but it also affects the network
of acquaintances.

In an earlier paper \cite{earlier} we introduced a model where
agents are represented by $N$ random walkers which diffuse on a square
$L\times L$ lattice and possibly interact if they are close together. This model
also takes into account the degradation of information when passing
from one agent to another: a decay constant $z$ quantifies such
alteration. As a consequence, the information spreading is a
history-dependent process which is governed by the rules underlying
the diffusion of $N$ random walkers on the given space.

Indeed, diffusion phenomena are dramatically affected by the
topology of the underlying space
\cite{burioni,polya,montroll,tauber,vanwijland}. For example, on
infinite lattices, the asymptotic probability for a random walker to
return to its starting point equals $1$ in one and two dimensions,
while for $d \geq 3$ there is a non-null probability to never return
to the starting point. The former case is said to be
\textit{recurrent}, while the latter is called \textit{transient}
\cite{burioni,polya,montroll}. Moreover, in many processes
concerning {\it interacting} random walkers, dimension 2 plays the
role of an upper critical dimension: it separates a
higher-dimensional regime where the mean-field results are exact
from a lower-dimensional regime where fluctuations become important.
This is the case, for example, for two-species diffusion-limited
reactions \cite{tauber}, or for the trapping of a random walker by
diffusing traps \cite{vanwijland}. In such processes, additional
logarithmic corrections for the power laws in dimension $d=2$
typically appear. Hence, in general, when dealing with diffusion one
should also wonder how the laws describing the problem are affected
by the geometry.

The model introduced in Ref. \cite{earlier} for $d=2$ is now
extended to $d$-dimensional hypercubic lattices. Numerical
simulations are carried for dimensions from $d=1$ up to $d=5$.
Analytical investigations are led which especially focus on the
one-dimensional case and on a mean-field approach which provides
good estimates for high-dimensional ($d \geq 3$) lattices. Most of
the results presented here do also hold for general Euclidean (i.e.,
translationally invariant) lattices, since the large-scale topology
of these systems (and quantities depending on it, such as the time
$\tau$ defined below in the low-density limit) depends solely on
their dimension, and not on small-scale details.

The main important quantities we are concerned with are the
Population-Awareness Time $\tau$, which represents the average time
necessary for the piece of information to reach the whole
population, and the final degree of information per agent
$\mathcal{I}_{ag}(z)$.

The time $\tau$ depends on the system parameters $N$ and $L$. Our
numerical results show that in the low-density regime, and in every
dimension $d$, this dependence can be factorized as
$\tau(N,L)=f(N)\,g(L)$, where $f$ and $g$ depend on $d$. Moreover,
in the low-density regime, we find the asymptotic behaviors of both
$f(N)$ and $g(L)$ to agree with mean-field calculations for $d \geq
3$, while for $d = 2$ deviations from the mean-field behavior appear
and for $d = 1$ the results are radically different. We therefore
argue that dimension $d=2$ is marginal for the phenomenon under
examination, and the mean field calculation of $\tau$ is exact for
$d > 2$.

The most important result contained in Ref. \cite{earlier} concerns
a nonasymptotic phenomenon: the nonmonotonic dependence of the final
degree of information per agent $\mathcal{I}_{ag}$ on $N$ and $L$,
with the emergence of extremal points. A process of optimization of
the final information is therefore intrinsically nontrivial. We show
here that the existence of extremal points is not a consequence of
the special choice $d=2$, but it arises in all dimensions $d\geq 1$.
It therefore appears as a {\it universal} and geometry-independent
phenomenon, occurring at the crossover between high- and low-density
regimes.

The paper is organized as follows. Section~\ref{sec:model} is
devoted to the description of the model.
Section~\ref{sec:Analytical} contains analytical results; it is
divided into high-density calculations (Sec.~\ref{sec:IIIA});
low-density calculations for $d=1$ (Sec.~\ref{sec:IIIB});
low-density calculations for $d>1$ (Sec.~\ref{sec:IIIC}).
Section~\ref{sec:NumRes} shows results obtained by means of
numerical simulations. We first consider the population awareness
time $\tau$ (Sec.~\ref{sec:num}), then the behavior of the final
degree of information and the quantities that affects it
(Secs.~\ref{sec:num_pop}, ~\ref{sec:FinInfo}). Finally,
Sec.~\ref{sec:Conclusions} includes our conclusions and
perspectives.

\section{\label{sec:model}The model}
The model analyzed in this work represents an extension of the one
introduced in an earlier paper \cite{earlier}. In this section we
briefly recall how it works.

We consider a population of $N$ random walkers (agents) moving on a
$d$-dimensional hypercubic lattice sized $L$ and endowed with
periodic boundary condition. Agents are initially ($t=0$)
distributed randomly throughout the whole volume $L^d$. We define
the density of agents as $\rho=N/L^d$; the ``low-density'' regime is
for $\rho\ll 1$ and the ``high-density'' regime for $\rho\gg 1$. At
each following instant each agent jumps randomly to one of the $2d$
nearest-neighbor sites. Notice that the same site can be occupied by
more agents, i.e, there are no excluded-volume effects.

At $t=0$ we assume that only one agent (called ``Information
Source'') carries information, while the remaining $N-1$ agents are
unaware. Two agents can then interact if their distance on the
underlying lattice is $\leq 1$ and if one of them is informed and
the other unaware. By ``interaction'' we mean an information passing
from the informed agent, say $j$, to the unaware one $k$ with a
fixed decay constant $z$ ($0\leq z\leq 1$): if $j$ carries
information $I_j$, then $k$ becomes informed with information
$I_k=z\,\cdot I_i$. Hence, the information carried by agent $j$ is
represented by the quantity $I_j$, $0\leq I_j \leq 1$, and, in
particular, when $I_j>0$ ($I_j=0$) the agent is ``aware''
(``unaware'').

Once an agent has become informed, it will never change nor lose its
information. As a consequence, there exists a final time
$t_{\mathrm{fin}}$ at which the total information of the system can
no longer evolve: at this time the information has reached every
agent and the simulation stops. Such time is a stochastic quantity
with average value $\tau$ called the Population-Awareness Time (PAT)
and standard deviation denoted as $\sigma_{\tau}$. Part of this work
is devoted to studying the properties of $\tau$ as a function of $L$
and $N$.

The total number of informed agents at a given time $t$ is again a
stochastic variable; we call $n(t)$ its average over all the
realizations of the system [$n(0)=1$; $n(\infty)=N$]. As a result of
our model, the information carried by an agent is always a power
$z^l$ of the decay constant, $l$ being the number of passages from
the Information Source to the agent. It is convenient to divide
informed agents into levels, so that an agent belongs to level $l$
when the information it receives has undergone $l$ passages from the
Information Source and equals $z^l$. We call $n(l,t)$ the number of
agents belonging to the $l$th level at time $t$, averaged over all
different realizations [$n(t)=\sum_{l=0}^{t}n(l,t)$]. The average
total information at time $t$ is therefore the generating function
of $n(t)$,
\begin{equation}
\label{eq:tot_information}
\mathcal{I}(z,\,t)=\sum_{l=0}^{t}n(l,t)z^l.
\end{equation}
In particular, we are interested in the final degree of information
$\mathcal{I}(z)$, that is the total information achieved once the
whole population has been informed,
\begin{equation}\label{eq:info_PAT}
\mathcal{I}(z) =\mathcal{I}(z,\infty)=\sum_{l=0}^{N}n(l,\infty)z^l.
\end{equation}
We also denote its average value per agent as
$\mathcal{I}_{ag}(z)=\mathcal{I}(z)/N$. The quantity $n(l,\infty)$
as a function of $l$ is called the final distribution of the
population on levels.

\section{\label{sec:Analytical} Analytical Results}

Although the model cannot be exactly solved in the general case, it
is possible to provide approximate solutions in some limit cases.
There are two time scales involved in the process: one for the
diffusion of the random walkers and one for the information passing.
When they are very different, approximate analytical approaches
become feasible and give results in good agreement with the
numerical simulations. When the two time scales are comparable, only
a numerical approach is possible (Sec.~\ref{sec:NumRes}).

In this Section we give analytical results for the PAT and the final
distribution on levels in two limit cases. Section~\ref{sec:IIIA}
treats the high-density ($\rho\gg1$) limit, with particular regard
to the case $d=1$. Section~\ref{sec:IIIB} considers the asymptotic
low-density ($\rho\ll1$) limit for $d=1$. A general mean-field
theory of this limit for all dimensions is given in
Sec.~\ref{sec:IIIC}.

\subsection{High-density regime}\label{sec:IIIA}

When $\rho\gg 1$, we can assume that the set of informed agents
covers a connected volume of the lattice, and that this volume
expands with a constant velocity (depending on the density $\rho$
and dimension $d$). We clarify this statement by considering $d=1$.

\begin{figure}
\includegraphics[width=0.45\textwidth]{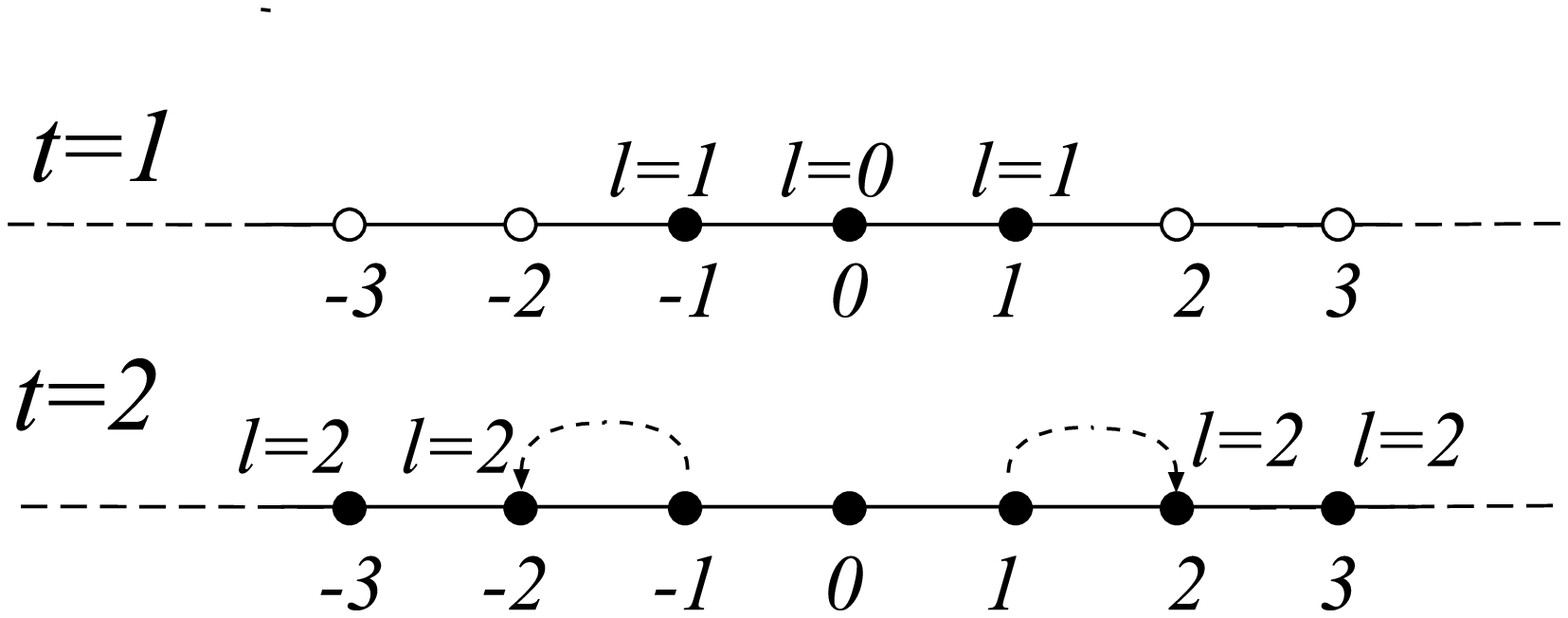}
\caption{\label{fig:highd_1d} High-density approximation for $d=1$.}
\end{figure}

Let us consider a chain of finite length $L$, with
$N\rightarrow\infty$ agents on it, and label the sites with the
numbers from 1 to $L$ (Fig. \ref{fig:highd_1d}). The number of
agents on a given site is $\rho$; for $N\rightarrow\infty$
($\rho\rightarrow\infty$), the probability that a given site is
empty is 0. Also, for $\rho\rightarrow\infty$, we assume that with
probability 1 at least one of the $\rho$ agents of a given site will
jump to the left {\it and} one to the right. Let the Source be in
$0$ at $t=1$: all the agents in $0$, $1$, and $-1$ will get
informed. At $t=2$ some of the newly informed agents jump on $\pm
2$; hence, the agents on $\pm 2$ and $\pm 3$ become informed. The
motion of the information front decouples from the random motion of
the agents; it expands with a {\it deterministic} law, with constant
velocity of 4 sites per time unit. The time required to cover the
whole chain is then
\begin{equation}\label{eq:tau_1D_ldil}\tau \simeq L/4.\end{equation}
The border of the informed zone contains all the newly-informed
agents, so each time step adds a new level and $n(l,\,\infty)=4\rho$
for $l\geq1 $. This is the origin, for high densities, of the
plateau observed in the one-dimensional distributions (Fig.
\ref{fig:Distr_1D_arrow}).

For $d>1$ it can be shown (see Ref. \cite{earlier} for the case
$d=2$) that the volume of the informed zone is a $d$-dimensional
polyhedron. The time it takes the border of the informed volume to
reach the border of the lattice is again $L/4$, but now $L/4$ more
instants are required to cover the rest of the lattice. Hence, in
this case
\begin{equation}\label{eq:tau_d>1_ldil}\tau \simeq L/2.\end{equation}
The new agents added at each step cover a $(d-1)$-dimensional
surface, hence $n(l,\infty)\simeq l^{d-1}$ for $l\leq L/4$, and
$n(l-L/4,\infty)\simeq n(L/4,\infty) - (l-L/4)^{d-1}$ for $l$ up to
$L/2$.

\begin{figure}
\includegraphics[width=0.45\textwidth]{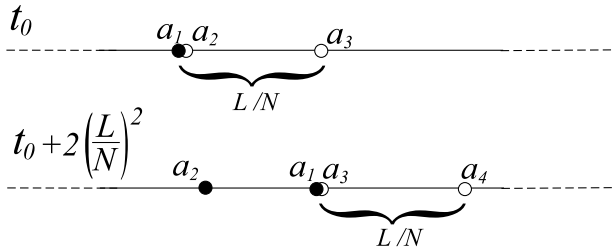}
\caption{\label{fig:lowd_1d} Low-density approximation for $d=1$.}
\end{figure}

\subsection{Low-density regime in $d=1$\label{sec:IIIB}}
Let us consider a chain of length $L$ and a population of $N$ agents
randomly distributed on it, with $N \ll L$ (Fig. \ref{fig:lowd_1d}).
The average distance among two agents is $\rho^{-1} = \frac{L}{N}$.
Due to the low-density hypothesis, we can neglect interactions
involving more than two agents. In this approach we divide the
problem of diffusion among $N$ agents into a sum of easier
(three-bodies) problems.

Let us consider the instant $t_0$ when the rightmost agent $a_1$
informs an unaware agent $a_2$ (Fig. \ref{fig:lowd_1d}). Let us call
$a_3$ the next unaware agent on the right: the average distance from
$a_3$ to $a_1$ and $a_2$ is $L/N=\rho^{-1}$ (if $\rho$ is small
enough, we can consider $a_1$ and $a_2$ to be on the same site). A
calculation regarding an epidemic model in one dimension similar to
ours \cite{warren} found that for low densities the velocity of the
front propagation approaches $\rho/2$. Hence, the average time it
takes one of the two aware agents to meet the unaware one is
$(L/N)/(\rho/2)=2(L/N)^2$. If we now suppose that $a_3$ has been
first reached by $a_2$, it will take again a time $2(L/N)^2$ for one
of them to reach the next unaware agent $a_4$ on the right, and so
on. There are about $N/2$ processes of this kind on the right-hand
side and $N/2$ on the left-hand side, which provides

\begin{equation}\label{eq:tau_1D}\tau \sim \frac{L^2}{N}.
\end{equation}

Now, if $a_1$ belongs to level $l$ ($a_2$ to level $l+1$), $a_3$
will belong to levels $l+1$ or $l+2$ with probabilities $1/2$; $a_4$
will belong to levels $l+1$, $l+2$ or $l+3$ with probabilities
$1/4$, $1/2$, and $1/4$, respectively, and so on. It is easy to show
by induction, starting from a Source on level 0, that at the $i$th
information passing the new agent on the left-hand side is on level
$l$ with probability $ 2^{-i}\left(\begin{array}{c}
i\\l\end{array}\right)$ ($0\leq l\leq i$); the same for the new
agent on the right-hand side. Hence, the average final number of
agents on level $l$ is
\begin{equation}\label{eq:info_1d}
n(l,\infty) \sim \sum_{i=0}^{N/2} 2^{-i}
\left(\begin{array}{c}i\\l\end{array} \right).
\end{equation}
There is no easy closed form for this sum, but it can be plotted for
any value of $N$: the curve displays a plateau of height 2, before
decaying to 0.

If we call $\Delta \mathcal{I} (t)$ the increment
of the total information at time $t$, we can write
\begin{equation}
\Delta \mathcal{I} (i+1) = \Delta \mathcal{I}(i) \frac{z+1}{2},
\end{equation}
and therefore
\begin{equation}\label{eq:info_1d_anal}
\mathcal{I}(z) =
\frac{z(z+1)}{2(1-z)}\left[1-\left(\frac{z+1}{2}\right)^{N/2}\right].
\end{equation}

\subsection{Low-density regime in $d>1$}\label{sec:IIIC}
In the case of low density
($\rho\ll 1$) the time an informed agent walks before meeting an
unaware agent becomes very large. We adopt a mean-field
approximation by assuming that the agents between each event have
the time to redistribute randomly on the lattice. In this
approximation, the probability that two given agents are in contact
at a given time is $p_d=\left\langle\tau_d\right\rangle^{-1}$, where
$\left\langle\tau_d\right\rangle$ is the average time for two random
walkers
to meet on a $d$-dimensional cubic lattice.\\
The process is an absorbing Markov chain with $N$ states; the system
is in state $k$ when it has $k$ informed agents. The chain starts
from state 1 and evolves to the absorbing state (state $N$). The
transition matrix $\mathbf{P}$ can be written: the transition
probability from a state $k$ to a state $m$ as a function of $N$ and
$p_d$ is
$$P_{k\,m}=
\left(
\begin{array}{c}
N-k\\
m-k
\end{array}
\right)
\left[1-\left(1-p_d\right)^k\right]^{m-k}\left[\left(1-p_d\right)^k\right]^{N-m}$$
for any $N$ and $p_d$. This is an upper triangular matrix, since the
binomial coefficient $\left(\begin{array}{c}
N-k\\m-k\end{array}\right)$ is 0 for $m<k$. We then make a
low-density approximation: we expand matrix $\mathbf{P}$ to first
order in $p_d$ to obtain
\begin{equation}
P_{k\,m} =
\begin{cases}
1 - m\,(N-m)\,p_d\;&\mathrm{for}\;m=k,\\\nonumber
m\,(N-m)\,p_d\;&\mathrm{for}\;m=k+1,\\\nonumber
0\;&\mathrm{elsewhere.}\end{cases}\nonumber
\end{equation}
This means that the system in the state $m$ has a probability $1 -
m\,(N-m)\,p_d$ to stay in $m$ and a probability $m\,(N-m)\,p_d$ to
jump to state $m+1$. We now take matrix $\mathbf{Q}$, the submatrix
obtained from $\mathbf{P}$ subtracting the last row and column
(those pertaining to the absorbing state), and compute the
fundamental matrix $\mathbf{F}=(1-\mathbf{Q})^{-1}$; a direct
calculation shows that $\mathbf{F}$ is an upper triangular matrix
given by
\begin{equation}
F_{k\,m} =
\begin{cases}
\frac{1}{m\,(N-m)\,p_d}, \;&\mathrm{for}\; k\geq m\\\nonumber
0&\mathrm{for}\; k < m.
\end{cases}\nonumber
\end{equation}

The mean time $\tau$ required to reach the absorbing state $N$
starting from state 1 is given by the sum of the first row of
$\mathbf{F}$,
\begin{equation}
\tau=\frac{1}{p_d}\sum_{m=1}^{N-1}\frac{1}{m(N-m)},
\end{equation}
and for $N\rightarrow\infty$,
\begin{equation}\label{eq:tau_d>3}
\tau\sim\frac{2}{N\,p_d}\left(\gamma+\mathrm{ln}(N)\right)=2\left\langle\tau_d\right\rangle \frac{\gamma+\mathrm{ln}(N)}{N},
\end{equation}
where $\gamma=0.577...$ is the Euler-Mascheroni constant.

A classical result \cite{montroll, traps} states that for
$d$-dimensional cubic lattices the asymptotic dependence of
$\left\langle\tau_d\right\rangle$ on the lattice size $L$ is
\begin{equation}
\left\langle\tau_d\right\rangle \sim
\begin{cases}
u_1\,L^2\;&d=1\\\nonumber u_2\,L^2\,\mathrm{ln}(L)&d=2\\\nonumber
u_d\,L^d&d>2,
\end{cases}\nonumber
\end{equation}
where the $u_d$ are dimension-dependent constants (for example,
$u_2=0.758...$). As we will show in the following, the asymptotic
dependence of $\tau$ on $L$ agrees with mean-field results for {\it
every} $d$, while the breakdown of the mean-field theory shows up in
the dependence on $N$ for $d\leq 2$.

It is possible to include the distribution on levels in the Markov
chain analysis, but the calculations become very lengthy and we give
only the final result. It is
\begin{equation}
n(l,\,\infty)=\frac{1}{(N-1)!}|s(N,\,l+1)|,
\end{equation}
and is exact for every $N$. Here, $|\cdots|$ denotes the absolute
value, and $s(m,\,k)$ is the Stirling number of the first kind
\cite{stirling}. The $s(m,\,k)$ are integers that appear in many
combinatorial problems; one of the possible asymptotic expansions
for Stirling numbers is \cite{wilf}
$$\frac{1}{(m-1)!}|s(m,k)|=\gamma\frac{\mathrm{ln}(m)^{k-1}}{(k-1)!}+O(\mathrm{ln}(m)^{k-2}),$$
so that
\begin{equation}\label{eq:hdil_lev}
n(l,\,\infty)\sim\frac{\mathrm{ln}(N)^{l}}{l!},
\end{equation}
which is the form we will use to fit the low-density distributions.
From this distribution it also follows that
\begin{equation}\label{eq:hdil_info}
\mathcal{I}\sim N^z.
\end{equation}

\section{\label{sec:NumRes}Numerical Results}

\subsection{\label{sec:num} Population-Awareness Time}

\begin{figure}[b]
\includegraphics[width=.40\textwidth]{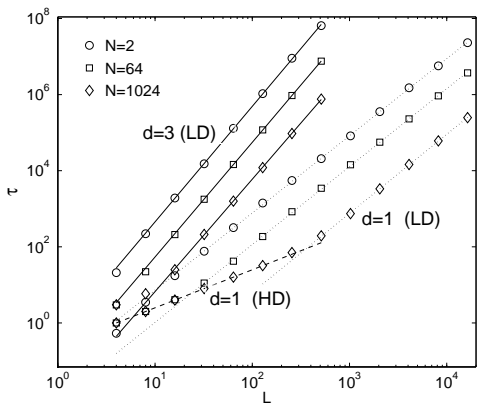}
\caption{\label{fig:Tau_L}Log-log scale plot of the
Population-Awareness Time $\tau$ versus the lattice size $L$.
Results obtained for a chain (dashed line) and for a cube (solid
line) are depicted. Different lattice-size values are shown with
different symbols, as explained by the legend. For large densities
($\rho \gg 1$, HD) and low densities ($\rho \ll 1$, LD), straight
lines represent the best fit according to Eq.~(\ref{eq:Tau_L_hd})
and Eqs. (~\ref{eq:Tau_L_1D}) and (\ref{eq:Tau_L_D}), respectively.
Error on data points is less than $2 \%$. The standard deviation is
not appreciable on this scale.}
\end{figure}

\begin{figure}[b]
\includegraphics[width=.40\textwidth]{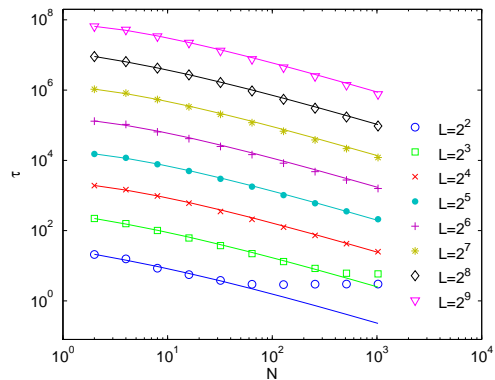}
\caption{\label{fig:Tau_N}(Color online) Dependence of the
Population-Awareness Time $\tau$ on the number of agents $N$ for the
cubic lattice $d=3$; different values of the lattice size $L$ are
shown with different symbols and colors. When the density of the
system is low, data points lie on the curve given by
Eq.~(\ref{eq:Tau_L_D}) which represents the best fit. Error on data
points is less than $2 \%$.}
\end{figure}

In this section we focus on numerical results concerning the
Population-Awareness Time $\tau$. We recall that $\tau$ has been
defined as the average time it takes the piece of information to
reach the whole population. Due to the analytical results discussed
in the preceding section, we expect the functional form displayed by
$\tau(N,L)$ to be strongly affected by the topology of the lattice
underlying the propagation.

Figures~\ref{fig:Tau_L} and ~\ref{fig:Tau_N} show the dependence of
$\tau$ on $L$ with $N$ fixed, and on $N$ with $L$ fixed,
respectively. In Fig.~\ref{fig:Tau_L}, where both the results for
$d=1$ and $d=3$ are displayed, two different regimes, of low and
high density, are clearly distinguishable for dimension $d=1$; for
$d=3$ the range of the high-density regime is too small, and only
the low-density one can be seen.

The high-density behavior is independent of $N$; indeed, for $\rho
\gg 1$ we find
\begin{eqnarray}\label{eq:Tau_L_hd}
&&\tau =\frac{L}{4}, \, d =1 \nonumber
\\
&&\tau =\frac{L}{2}, \, d > 1,
\end{eqnarray}
in agreement with Eqs. (\ref{eq:tau_1D_ldil}) and
(\ref{eq:tau_d>1_ldil}).

In the low-density regime ($\rho \ll 1$) and for $d\neq2$, $\tau$
follows the behavior calculated in Secs.~\ref{sec:IIIB}
and~\ref{sec:IIIC}. For $d=1$ the results as a function of $L$ and
$N$ are fitted by
\begin{equation}\label{eq:Tau_L_1D}
\tau \sim C_1N^{\alpha}L^{\beta},
\end{equation}
with $C_1=0.96(5)$; $\alpha=-0.98(5)$; $\beta=1.98(2)$, in agreement
with Eq.(\ref{eq:tau_1D}).

For $d=3,\,4$ we found the best fit for $\tau$ to be given by the
function
\begin{equation}\label{eq:Tau_L_D}
\tau \sim C_d\frac{\mathrm{ln}(L)+A}{N}L^{\beta},
\end{equation}
where $\beta=d$ within a $1\%$ error and $A = 0.59(3)$, in agreement
with the value $\gamma=0.577\ldots$ found in Eq. (\ref{eq:tau_d>3}).
The values of the constants are $C_3=0.77(1)$; $C_4=0.39(1)$
(different in general from the $u_d$ of the mean-field
approximation).

The low-density limit in the case $d=2$ deserves a separate
discussion. It is still possible to express $\tau$ as a product of
two distinct functions,
\begin{equation}
\tau \sim f_2(N)\,L^2\, \mathrm{ln}(L), \;\; \mathrm{for} \; d=2.
\end{equation}
The dependence on $L$ is in agreement with the improved mean-field
calculation (see Sec.~\ref{sec:IIIC}). This best fit is better than
that in Ref. \cite{earlier}, where we hypothesized a non-integer
power law ($L^{2.2}$).

The analytical form of $f_2(N)$ cannot be unequivocally determined
by the simulations. In the fitting range the function $\frac{A +
\mathrm{ln}(N)}{N}$ agrees with the numerical results better than
the power-law $N^{-0.66}$ previously given \cite{earlier}. However,
in this case the value of the fitting parameter is $A=-0.18(2)$,
hence is definitely different from the mean-field value $\gamma$.
Since there are no analytical calculations to support this
functional form with this particular value of the fitting constant
for $d=2$, we cannot rule out higher-order logarithmic corrections.

To summarize, in the low-density regime, the function $\tau(N,L)$
factorizes into two parts, depending, respectively, on $L$ and $N$:
\begin{equation}\label{eq:Tau}
\tau \sim \ \left\{
\begin{array}{ccc}
\displaystyle  C_1\frac{L^2}{N}, & d=1, \\
\\
f_2(N)\,L^2\, \mathrm{ln}(L), & d=2, \\
\\
\displaystyle  C_d\frac{\gamma + \mathrm{ln}(N)}{N}\,L^d, & d\geq3,
\end{array}
\right.
\end{equation}
the $C_d$ being dimension-depending constants. The most satisfying
fitting function we have found for $f_2(N)$ is $\frac{A +
\ln(N)}{N}$, $A\simeq -0.18$.

The standard deviation $\sigma_{\tau}$ displays a similar dependence
on $N$ and $L$ for low densities: $\sigma_{\tau}\varpropto
N^{-1}L^2$ for $d=1$, and so on. For high densities $\sigma_{\tau}$
becomes vanishingly small, which is explained by the fact that the
propagation of information becomes a deterministic process.

\subsection{\label{sec:num_pop} Final Distribution on Levels: Universality
of the Extremal Distribution}

\begin{figure}[t]
\includegraphics[width=.5\textwidth]{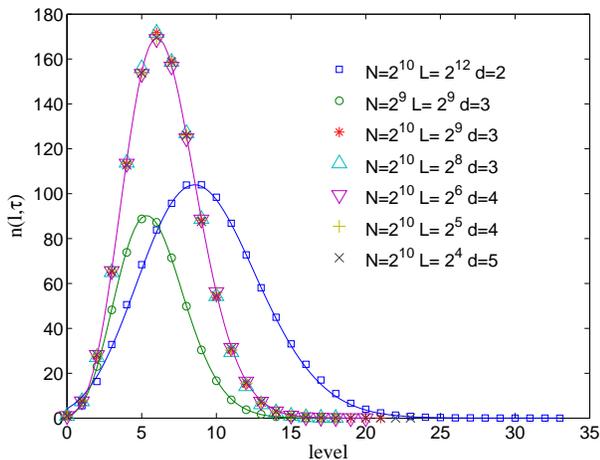}
\caption{\label{fig:Distr_3-4-5}  (Color online) Final population
distribution on levels $n(l,\,\infty)$ for low-density systems. Data
points agree with the fitting line drawn according to Eq.
(\ref{eq:Distr_3D}). For $d=2$ the fitting parameters depend
smoothly on both $L$ and $N$, and the curve is distinct from those
of higher dimension and same $N$. For $d\geq 3$, systems of
different dimension $d$ and size $L$ display distributions that
overlap within the error. Only the dependence on $N$ is left, as is
shown for $d=3$, $N=512$.}
\end{figure}

In Sec.~\ref{sec:model} we introduced the function $n(l,\,\infty)$,
which represents the final distribution of agents on levels and is
strongly connected with the final degree of information
$\mathcal{I}(z)$. The asymmetrical-bell shape displayed by the
distributions for hypercubic lattices with dimension $d>2$
(Fig.~\ref{fig:Distr_3-4-5}) and the way they evolve while varying
the system parameters $N$ and $L$ are analogous to the
two-dimensional case \cite{earlier}.

\begin{figure}
\includegraphics[width=0.4\textwidth]{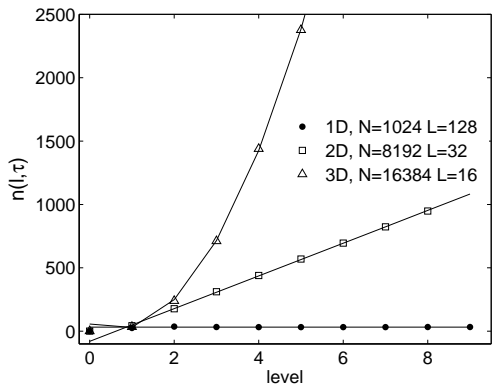}
\caption{\label{fig:distr_laws } Final population distribution on
levels for high densities and $d=1,2,3$. The dependence on $l$ is a
power law: $n(l,\,\infty)\sim l^{d-1}$.}
\end{figure}

In the limit case $\rho\gg 1$ (Fig. \ref{fig:distr_laws }) and in
every dimension, the final distribution on levels follows the law
$n(l,\,\infty)\sim l^{d-1}$, in agreement with the calculation in
Sec.~\ref{sec:IIIA}.

For $\rho$ small enough (i.e., for $L > \tilde{L}$ and $N <
\tilde{N}$, see below) the population distribution on levels for
$d\geq 2$ is well fitted by the following function:
\begin{equation}\label{eq:Distr_3D}
\frac{n(l,\infty)}{N}=A\,\frac{\left(\ln\,N\right)^l}{\Gamma(B\,
l+C)},
\end{equation}
where $\Gamma(x)$ is the Euler gamma function
(Fig.~\ref{fig:Distr_3-4-5}). The previous equation is a
generalization of Eq. (\ref{eq:hdil_lev}) found in the mean-field
approximation for the low-density regime.

The fitting parameters $A, B, C$ at low densities for $d=2$ are
smoothly dependent on $N$ and $L$, while those for $d>2$ are all
close to $1$ and independent of the lattice size $L$, keeping only
the dependence on $N$. Moreover, the data points for $d \geq 3$ with
the same values of $N$ {\it all collapse on the same curve}. This
holds for all the regular lattices with $d \geq 3$ we have
considered: the distribution curves at low densities are independent
of $d$, and agree with the mean-field form [Eq. (\ref{eq:hdil_lev}].

\begin{figure}[t]
\includegraphics[width=.5\textwidth]{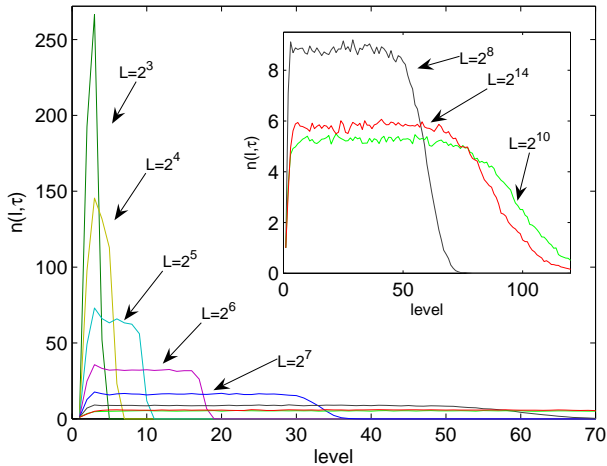}
\caption{\label{fig:Distr_1D_arrow}(Color online) Population
distribution on levels at $t=\tau$ for one-dimensional systems with
$N=512$ and $L$ ranging from $2^3$ to $2^{14}$, as shown by the
legend (the lines are guides to the eye). The behavior of the
distribution is nonmonotonic with respect to $L$: by increasing $L$
from small values, the curves first shift to the right and flatten;
the rightmost, extremal curve corresponds to $L=1024$. Then, by
further increasing $L$, the curves shift back to the left and
sharpen. The inset shows in detail the shift back with the curves
pertaining to $L=2^8, L=2^{10}, L=2^{14}$.}
\end{figure}

The description given so far concerns $d\geq2$ lattices; in
Fig.~\ref{fig:Distr_1D_arrow} we show for $N=512$ and varying $L$
the one-dimensional case, which exhibits quite different
distributions. Such distributions are still very sharp for very high
values of the density $\rho$, but soon develop a plateau by
increasing $L$; the plateau persists up to low densities. The
existence of a plateau was justified both in a high-density and in a
low-density approximation (Sec.~\ref{sec:Analytical}).

The main result of Ref. \cite{earlier} concerned the existence of an
\textit{extremal} curve for the distribution of agents on levels. We
have found that this feature does not depend on the dimension $d$ of
the lattice. We show in Fig.~\ref{fig:Distr_1D_arrow} how the
extremal distribution emerges in $d=1$, as a function of $L$, for a
particular value $L=\tilde{L}$ (here, $\tilde{L}\simeq 1024$), and
keeping $N$ fixed, notwithstanding the fact that its shape is
dramatically different with respect to the higher-dimensional ones.
While $L<\tilde{L}$, the distribution displays a plateau whose
height (width) is a monotonically decreasing (increasing) function
of the chain length $L$; the distribution curve shifts to the right.
Conversely, for $L>\tilde{L}$ a shift-back phenomenon analogous to
that discussed in Ref. \cite{earlier}: now, by rising $L$, the
height gets larger while the width gets smaller. As can be seen from
Fig. \ref{fig:Tau_L}, $\tilde{L}$ corresponds to the crossover
between high- and low-density regimes. The same happens by varying
$N$ and keeping $L$ fixed; there is an extremal distribution for a
particular value $\tilde{N}$, depending on $L$, and corresponding to
the crossover between the two regimes.

This shift-back phenomenon, and the existence of an extremal
distribution, occur in all the dimensions we have investigated (up
to $d=5$). It therefore constitutes a universal feature,
independent of lattice dimension, and, as we will see, it provides
striking effects on the final degree of information.

\subsection{\label{sec:FinInfo} Degree of Information}
In this section we deal with the final degree of information
$\mathcal{I}(z) = \mathcal{I}(z,\infty)$ (\ref{eq:info_PAT}) and its
dependence on the decay constant $z$ and system parameters $N$, $L$.
We remind [Eq. (\ref{eq:tot_information})] that $\mathcal{I}(z)$ is
the generating function of the final populations $n(l,\,\infty)$,
hence its value depends on the final distribution of the population
on levels analyzed in the preceding paragraphs.

\begin{figure}[b]
\includegraphics[width=.4\textwidth]{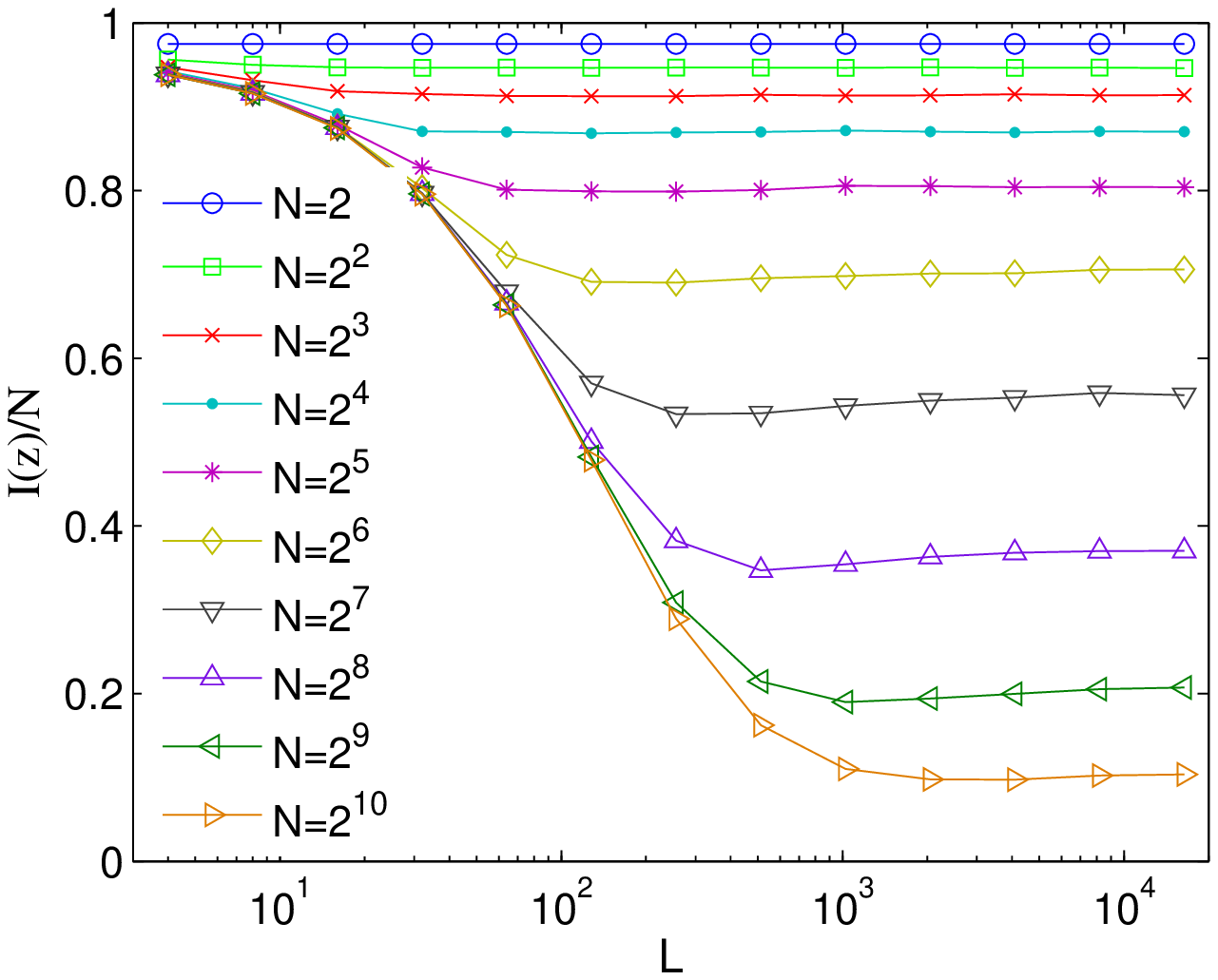}
\includegraphics[width=.4\textwidth]{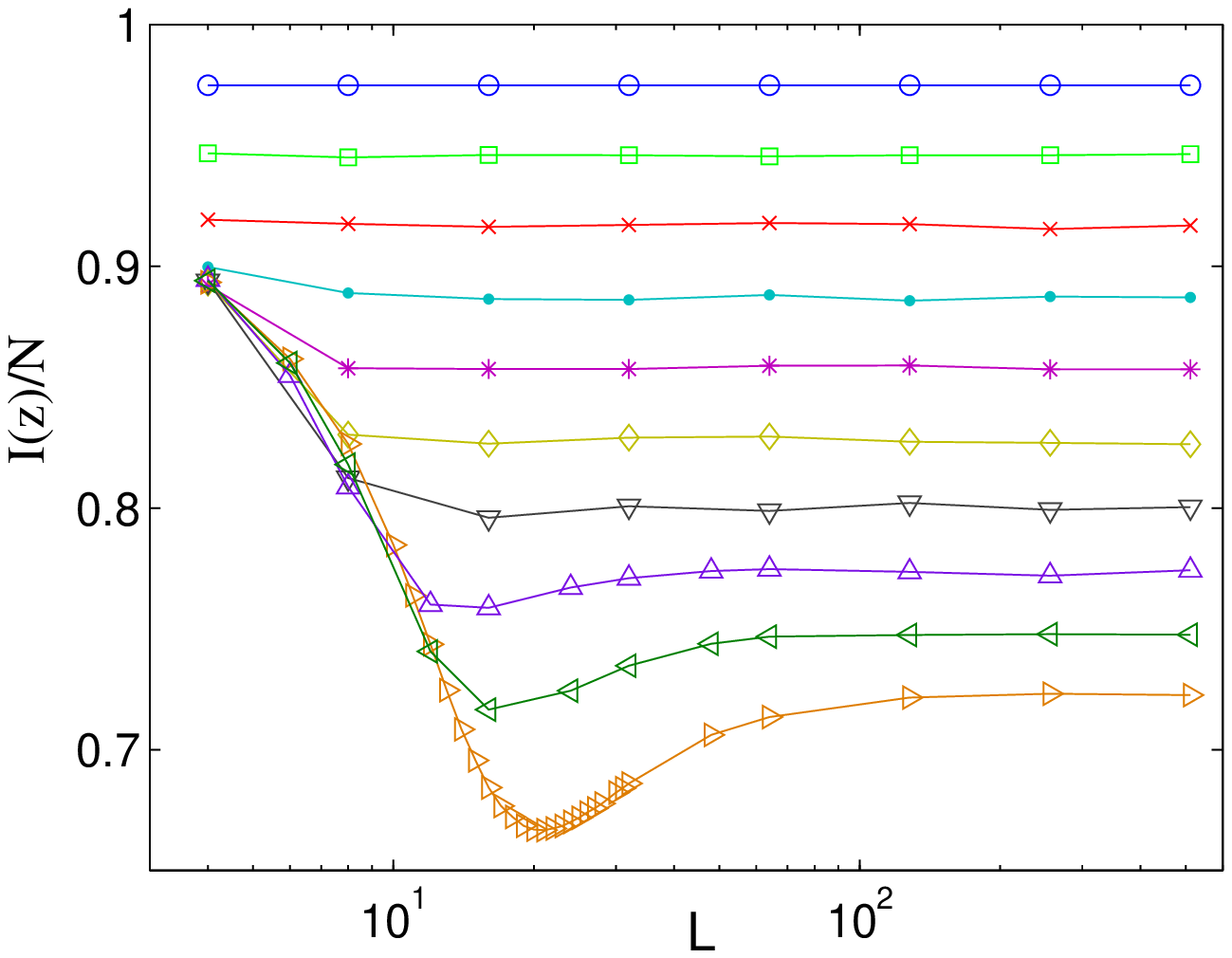}
\caption{\label{fig:Info}  (Color online) Semilog scale plot of
final degree of information per agent
$\mathcal{I}_{ag}(z)=\mathcal{I}(z)/N$ vs lattice size $L$, for
$d=1$ (top) and $d=3$ (bottom). The decay constant is fixed at
$z=0.95$. Several values of $N$ are shown with different symbols and
colors (lines are guides to the eye) and the legend is the same for
both figures. Notice that minimum depth is greater for the latter
case. Error on data points is $<4 \%$.}
\end{figure}

Let us firstly consider the dependence on the decay constant $z$.
Again, results highlight strong differences between the
one-dimensional and higher-dimensional cases ($d \geq 2$). In the
latter case and for the low-density regime (approximately $\rho <
2^{-8}$), we find
\begin{equation}\label{eq:info_ld}
\mathcal{I}(z) = N^{z},
\end{equation}
within the error ($ <4 \% $).

On the other hand, when $d=1$, the final degree of information
shows an exponential growth which can be represented by the
following equation:
\begin{equation}\label{eq:info_ld_1D}
\mathcal{I}(z) =A \frac{z(1+z)}{1-z}(1-e^{-B \cdot N(1-z)}),
\end{equation}
where $A$ and $B$ smoothly depend on $N$ and $L$. Equations
(\ref{eq:info_ld}) and (\ref{eq:info_ld_1D}) are in very good
agreement with the expressions found in the mean-field approximation
[Eqs. (\ref{eq:hdil_info}) and (\ref{eq:info_1d_anal}),
respectively].

Once $z$ is fixed, $\mathcal{I}_{ag}(z)$ depends nonmonotonically on
$N$ and $L$: let us follow it for $N$ fixed and varying $L$ in
Fig.~\ref{fig:Info} in the two cases $d=1$ and $d=3$. For $L$ small,
due to the narrow distribution discussed in the preceding section,
the value of the information is high. When $L=\tilde{L}$, the
population distribution on levels reaches its extremal form and the
information displays a minimum. As $L$ increases, the information
starts to rise again (as can be seen, the effect gets more marked by
increasing the dimension). Hence, given a population number $N$,
there is an optimal lattice size $\tilde{L}$ for which the final
information is minimum. The same happens having fixed $L$ and
letting $N$ vary: there is a minimum for $N=\tilde{N}$, depending on
$L$. As underlined in Ref. \cite{earlier}, the existence of a local
minimum of the final information implies that choosing an
optimization strategy for the spreading of information on the
lattice is not trivial. There is no {\it a priori} right direction
in parameter space where to move in order to improve
$\mathcal{I}(z)$; rather, the direction depends on the starting
point.

\section{\label{sec:Conclusions} Conclusions and perspectives}

In this work the model of information spreading previously
introduced has been extended to different geometries; indeed, we
considered the chain and $d$-dimensional hypercubic lattices. The
occurrence of a nonmonotonic behavior for the final degree of
information is not due to a special geometry underlying the process,
but its origin lies in the crossover between the two different
regimes of high and low density. Therefore, the existence of minima
in the final degree of information is universal and, remarkably,
even the possibility to derive optimization strategies does not
depend on the particular structure the process is embedded in.

On the other hand, the asymptotic laws for $\tau$ are interestingly
related to the geometry underlying the random-walk diffusion. In
particular, $d=2$ is a marginal dimension separating two
well-behaved cases, which suggests an investigation on in-between
dimensions \cite{future}.

The robustness of the existence of extremal point for $\mathcal{I}$
is an important point since the possibility of extracting optimal
strategies is not a feature restricted to some special structures.


\begin{thebibliography}{}
\bibitem{traps}
Indeed, the result in \cite{montroll} concerns the mean absorption
time into a trap at the origin for a single walker. However, the
problem of the mean first meeting time of two walkers, 1 and 2, can
be mapped onto the problem of absorption at the origin with a proper
rescaling of time.
\end{thebibliography}
\end{document}